\documentclass[twocolumn,letterpaper]{IEEEAerospaceCLS}  

\usepackage[]{graphicx}    
\newcommand{\ignore}[1]{}  

\usepackage{mathrsfs}

\usepackage{psfrag,epsfig,graphics}
\usepackage{amsmath,amssymb,multirow}

\DeclareSymbolFontAlphabet{\amsmathbb}{AMSb}%

\usepackage{booktabs}
\usepackage{graphicx,amsmath,graphicx}
\usepackage{caption}
\usepackage[position=b]{subcaption}

\usepackage[noadjust]{cite}
\usepackage{multirow}
\usepackage[lined,linesnumbered,ruled]{algorithm2e}

\usepackage{color}  

\newcommand{\cb}[1]{{\boldsymbol{#1}}}
\newcommand{\cp}[1]{\ifmmode {\mathcal{#1}}\else ${\mathcal{#1}}$\fi}

\newcommand{\bD}{\boldsymbol{D}}

\newcommand{\bF}{\boldsymbol{F}}
\newcommand{\bG}{\boldsymbol{G}}
\newcommand{\bH}{\boldsymbol{H}}
\newcommand{\bI}{\boldsymbol{I}}

\newcommand{\bK}{\boldsymbol{K}}

\newcommand{\bP}{\boldsymbol{P}}
\newcommand{\bQ}{\boldsymbol{Q}}
\newcommand{\bR}{\boldsymbol{R}}
\newcommand{\bS}{\boldsymbol{S}}

\newcommand{\bT}{\boldsymbol{T}}

\newcommand{\bc}{\boldsymbol{c}}

\newcommand{\bq}{\boldsymbol{q}}
\newcommand{\br}{\boldsymbol{r}}
\newcommand{\by}{\boldsymbol{y}}
\newcommand{\bs}{\boldsymbol{s}}

\newcommand{\bv}{\boldsymbol{v}}
\newcommand{\bx}{\boldsymbol{x}}

\newcommand{\bz}{\boldsymbol{z}}

\newcommand{\calD}{\mathcal{D}}

\newcommand{\calH}{\mathcal{H}}

\newcommand{\calN}{\mathcal{N}}

\newcommand{\bbZ}{\mathbb{Z}}

\usepackage[euler]{textgreek}
\usepackage[colorinlistoftodos]{todonotes}

\newcommand{\bmu}{\boldsymbol{\mu}}

\newcommand{\bSigma}{\boldsymbol{\Sigma}}

\newcommand{\vect}{\operatorname{vec}}

\usepackage{color}  

\definecolor{darkgreen}{rgb}{0., 0.4, 0.}
\definecolor{amber}{rgb}{1.0, 0.49, 0.0}
\definecolor{orange}{rgb}{1.0, 0.4, 0.0}
\definecolor{darkorange}{rgb}{0.7, 0.32, 0.0}

\renewcommand{\calH}{\mathcal{H}}
\renewcommand{\calD}{\mathcal{D}}

\begin{document}
\title{Online Multi-resolution Fusion of Space-borne Multispectral Images}


\author{%
Haoqing Li\\
Department of ECE\\
Northeastern University\\
Boston, MA 02115\\
li.haoq@northeastern.edu
\and
Bhavya Duvvuri\\
Department of CEE\\
Northeastern University\\
Boston, MA 02115\\
duvvuri.b@northeastern.edu
\and
Ricardo Borsoi\\
Department of EE\\
Federal University of Santa Catarina\\
Florianópolis, SC, Brazil\\
raborsoi@gmail.com
\and
Tales Imbiriba\\
Department of ECE\\
Northeastern University\\
Boston, MA 02115\\
talesim@ece.neu.edu
\and
Edward Beighley\\
Department of CEE\\
Northeastern University\\
Boston, MA 02115\\
r.beighley@northeastern.edu
\and
Deniz Erdo{\u{g}}mu{\c{s}}\\
Department of ECE\\
Northeastern University\\
Boston, MA 02115\\
erdogmus@ece.neu.edu
\and
Pau Closas\\
Department of ECE\\
Northeastern University\\
Boston, MA 02115\\
closas@ece.neu.edu
%
\thanks{\footnotesize 978-1-6654-3760-8/22/$\$31.00$ \copyright2022 IEEE}
}

\maketitle

\thispagestyle{plain}
\pagestyle{plain}

\maketitle

\thispagestyle{plain}
\pagestyle{plain}

\begin{abstract}
Satellite imaging has a central role in monitoring, detecting and estimating the intensity of key natural phenomena. One important feature of satellite images is the trade-off between spatial/spectral resolution and their revisiting time, a consequence of design and physical constraints imposed by satellite orbit among other technical limitations. In this paper, we focus on fusing multi-temporal, multi-spectral images where data acquired from different instruments with different spatial resolutions is used. We leverage the spatial relationship between images at multiple modalities to generate high-resolution image sequences at higher revisiting rates. To achieve this goal, we formulate the fusion method as a recursive state estimation problem and study its performance in filtering and smoothing contexts. The proposed strategy clearly outperforms competing methodologies, which is shown in the paper for real data acquired by the Landsat and MODIS instruments.
\end{abstract}

\tableofcontents

\section{Introduction}

High spatial resolution satellite image data is a fundamental tool for remote sensing applications such as the monitoring of land cover changes~\cite{lu2016land, zhu2014continuous}, deforestation~\cite{portillo2012forest,schultz2016performance} or water mapping~\cite{kim2017mapping, yoon2016estimating} and water quality~\cite{gholizadeh2016comprehensive}.
Moreover, to adequately deal with the variability of such events over time it is important to have short time spans between different image acquisitions of the same scene (i.e., a high temporal resolution, or low revisit times).
However, fundamental limitations of multiband imaging instruments and large sensor-to-target distances impose a trade-off between spatial and temporal resolutions of satellite image sequences.

This means that instruments providing high spatial resolution have long revisit times, while the converse holds for instruments with short revisit times.
This can be illustrated, for instance, by considering Landsat~8 and MODIS instruments (with 30 and 250/500 meters spatial resolution, respectively). While MODIS is able to provide daily images at coarse resolution, Landsat-8 only revisits the same site once every 16 days~\cite{roy2014landsat}.

Considering these limitations, many works proposed multimodal image fusion techniques to generate high (spatial, spectral or temporal) resolution remote sensing images.
Multimodal image fusion aims to combine multiple observed images, each of which having high resolution in a given dimension -- spatial, temporal, or spectral -- to generate high resolution image sequences.

Fusing images with different spectral and spatial resolutions has been extensively studied to generate images with high spatial and spectral resolutions, which are critical for accurately distinguishing different materials in a pixel~\cite{yokoya2017HS_MS_fusinoRev,Borsoi_2018_Fusion,loncan2015pansharpeningReview}. Recently, an increasing interest has been observed in applying multimodal image fusion to generate image sequences with high spatial and temporal resolutions~\cite{belgiu2019spatiotemporalFusionRev}.
Existing spatiotemporal image fusion methods are usually divided in weighted fusion, umixing-based, learning-based and Bayesian approaches~\cite{zhu2018spatiotemporalFusReview}. There also exist hybrid techniques, which leverage ideas from more than one family of approaches.


Weighted fusion methods assume that the temporal changes occurring between two time instants are consistent between the high and low spatial resolution images for low resolution pixels which are composed of only a single material~\cite{gao2015fusingLandsatMODISreview}. However, coarse resolution pixels are often mixtures of different materials. The predicted high resolution pixels are then computed as a weighted linear combination of the previous high resolution pixels and of the changes occurring at low resolution pixels in a given neighborhood~\cite{gao2006STARFM,zhu2010fusion_ESTARFM}. Different works have designed various weighting functions, which aim to select neighboring pixels that are homogeneous and spatially/spectrally similar to the pixel whose change is being predicted~\cite{gao2006STARFM,zhu2018spatiotemporalFusReview}.
Other works have extended such framework account for sudden changes~\cite{hilker2009STAARCH_fusion} or to use different weighting functions~\cite{zhu2010fusion_ESTARFM}.


Unmixing-based methods make use of the linear mixing model (LMM), which assumes that each pixel in the low resolution image can be represented as a convex combination of the reflectance of a small number of pure spectral signatures, called endmembers~\cite{keshava2002unmixingReview,li2021AEC_SU_modelbased,borsoi2020BMUAN}.
The LMM has been used for multimodal image fusion by assuming the proportions of each material in a low resolution pixel to be stable/constant over time~\cite{zurita2008unmixingFusion1,amoros2013multitemporalFusionUnmixing,wu2012unmixingFusion2}. This way, spectral unmixing~\cite{keshava2002unmixingReview} is used to estimate the endmembers at different time instants from low resolution images, while using different strategies to mitigate the spectral variability of a single material~\cite{zurita2008unmixingFusion1,borsoi2020variabilityReview,Borsoi_multiscaleVar_2018}. However, abrupt abundance variations are commonly found in multitemporal image streams~\cite{liu2019reviewCD,borsoi2021MT_MESMA,erturk2015sparseSU_CD}, which may negatively impact the performance of such methods.


Learning-based approaches leverage training data and different machine learning algorithms in order to perform image fusion. Such approaches are varied, ranging from model-based approaches which use dictionary learning~\cite{huang2012spatiotemporalFusionDictLearning} (often based on sparse representation of image pixels by the LMM~\cite{borsoi2018superpixels1_sparseU}), to more flexible but model-agnostic methods such as convolutional neural networks~\cite{song2018spatiotemporalFusionCNNs}.


Bayesian methods are flexible alternatives to the previous approaches that take into account the uncertainty present both in the imaging model and in the estimated images. 
The Bayesian framework is based on the definition of probabilistic models to describe the relationship between images of different spatial, spectral and temporal resolutions acquired by different instruments. This allows image fusion to be formulated as a maximum a posteriori estimation problem~\cite{shen2016integratedSpatioTemporalFusion}.
Although Bayesian methods usually consider Gaussian distributions for mathematical tractability, different variations have been proposed depending on how the image acquisition process is modelled and on how the mean and covariance matrices are estimated. This included assuming them diagonal~\cite{huang2013spatiotemporalFusionBayes}, estimating image covariance matrices based on an initial estimate of the high resolution image~\cite{shen2016integratedSpatioTemporalFusion}, or based on the low resolution image pixels~\cite{xue2017bayesianFusionPixelCovariances}.

A recent work considered a Kalman filter-based approach to estimate a high resolution image sequence based on mixed resolution observations from the Landsat and MODIS instruments~\cite{zhou2020kalmanFusionLandsatMODIS}. However, to define the model for the Kalman filter, two Landsat+MODIS image pairs at times $t_0$ and $t_N$ are considered, as well as a time series of MODIS images at instants $t_k\in[t_0,t_N]$, making it unsuitable for online operation. Moreover, changes between each pair of images were assumed to be constant/uniform over predefined groups of high resolution image pixels, which can be restrictive (due to the large resolution difference between the measured images, the groups must contain many pixels in order to make the model well-posed). It also does not benefit from auxiliary information that could aid the estimation of the high resolution images.
Another work used the Kalman filter to estimate normalized difference vegetation indices (NDVI) time series images from Landsat and MODIS observations, using an affine model for the dynamics of the states whose coefficients are selected based on the seasonality, and another affine model to relate the NVDI estimate obtained from MODIS and Landsat measurements~\cite{sedano2014kalmanFUsionNDVI}.

In this paper, we propose a Kalman filter and smoother framework for spatio-temporal fusion of multispectral images. Unlike previous approaches, the proposed method is not restricted in the image modalities which must be observed at each time instant, and is able to operate online. We also develop a smoother to optimally exploit information contained in future observed images when processing images in a time window. Moreover, instead of considering the changes to be constant over areas comprising large amounts of image pixels, we propose a classification-based strategy 
to define a more informative time-varying dynamical image model by leveraging historical data. This allows for a better localization of changes in the high resolution image even in intervals where only coarse resolution observations (e.g., MODIS) are available.  
We illustrate the application of the proposed framework by fusing images from the Landsat and MODIS instruments. Experimental results indicate that the proposed method can lead to considerable improvements compared to using a non-informative dynamical model and to widely used image fusion algorithms, both in image reconstruction and in downstream water classification and hydrograph estimation tasks.

This paper is organized as follows. In Section~\ref{sec:model}, we present the paper notation and the proposed imaging model. Section~\ref{sec:filter} presents the Kalman filter and smoother approaches for multimodal image fusion. Section~\ref{sec:experiments} contains simulation experiments that illustrate the performance of the proposed method. Finally, Section~\ref{sec:conclusions} concludes the paper.










\section{Dynamical Imaging Model}
\label{sec:model}

\subsection{Definitions and notation}

Let us denote the the $\ell$-th band of the $k$-th acquired image reflectances from modality $m\in\Omega$ by $\by_{k,\ell}^{m}\in{\amsmathbb{R}}^{N_{m,\ell}}$, with $N_{m,\ell}$ pixels for each of the bands $\ell=1,\ldots,L_m$, and $\Omega$ denoting the set of image modalities. 
As a practical example, we consider $\Omega=\{\mathsf{L},\mathsf{M}\}$ to contain the Landsat-8, and MODIS image modalities, without loss of generality. 
We also denote by $\Omega_H$ the highest resolution image modality, e.g., $\Omega_H=\{\mathsf{L}\}$.
%
%
We denote the corresponding high resolution latent reflectances by $\bS_k\in{\amsmathbb{R}}^{N_H\times L_H}$, with $N_H$ pixels and $L_H$ bands, with $L_H\geq L_m$ and $N_H\geq N_{m,\ell}$, $\forall \ell,m$. Subindex $k\in{\amsmathbb{N}}_*$ denotes the acquisition time index. We also denote by $\operatorname{vec}(\cdot)$, $\operatorname{col}\{\cdot\}$, $\operatorname{diag}\{\cdot\}$ and by $\operatorname{blkdiag}\{\cdot\}$ the vectorization, vector stacking, diagonal and block diagonal matrix operators, respectively. The notation $\bx_{a:b}$ for $a,b\in{\amsmathbb{N}}_*$ represents the set $\{\bx_a,\bx_{a+1},\ldots,\bx_b\}$. We use ${\calN}(\bmu,\bSigma)$ to denote a Gaussian distribution with mean $\bmu$ and covariance matrix $\bSigma$.

\subsection{Measurement model}
To formulate our measurement model we assume that the acquired image at time index $k$, for any imaging modality, is a spatially degraded and spectrally transformed version of the high resolution latent reflectance image $\bS_k$.
Following this assumption our measurement model for the $m$-th modality becomes:
\begin{align}
    \by_{k,\ell}^{m} = {\calH}_{\ell}^{m}(\bS_k)\bc_{\ell}^{m} + \br_{k,\ell}^{m} \,, \,\,\, \ell=1,\ldots,L_m \,,
    \label{eq:obs_mdl_1}
\end{align}
where $\bc_{\ell}^{m}\in{\amsmathbb{R}}^{L_H}$ denotes a spectral transformation vector, mapping all bands in $\bS_k$ to the $\ell$-th measured band at modality $m$; ${\calH}_{\ell}^{m}$ is a linear operator representing the band-wise spatial degradation, modeling blurring and downsampling effects of each high resolution band, and $\br_{k,\ell}^{m}$ represents the measurement noise. Note that, while we consider the spatial resolution of the high resolution bands in $\bS_k$ to be the same, different bands from the same modality can have different resolutions. We also assume the measurement noise to be Gaussian and uncorrelated among bands, that is, $\br_{k,\ell}^{m}\sim{\calN}(\cb{0},\bR_{\ell}^m)$ with time-invariant covariance matrix given by $\bR_{\ell}^m\in{\amsmathbb{R}}^{N_{m,\ell}\times N_{m,\ell}}$, and $\operatorname{cov}(\br_{k,j}^{m},\br_{k,\ell}^{m})=\cb{0}$ for all $j\neq\ell$.
%

Note that satellite images may be corrupted by several effects, including dead pixels in the sensor, incorrect atmospheric compensation, and the presence of heavy cloud cover. Such pixels cannot be reliably used in the image fusion process as they may degrade the performance of the method. Directly addressing these effects using a statistical model would require the choice of a non-Gaussian distribution for the noise vector $\br_{k,\ell}^{m}$, which could make the computational complexity of the fusion procedure prohibitive.
Thus, we consider a matrix $\bD_k^{m}\in{\amsmathbb{R}}^{\widetilde{N}_m\times N_m}$, which eliminates outlier pixels from the image, leading to the following transformed measurement model:
\begin{align}
    \widetilde{\by}_{k,\ell}^{m} =  \bD_k^{m}{\calH}_{\ell}^{m}(\bS_k)\bc_{\ell}^{m} + \widetilde{\br}_{k,\ell}^{m} \,,
    \label{eq:obs_mdl_2}
\end{align}
where $\widetilde{\by}_{k,\ell}^{m}=\bD_k^{m}\by_{k,\ell}^{m}$ and $\widetilde{\by}_{k,\ell}^{m}=\bD_k^{m}\br_{k,\ell}^{m}$ denotes the measured image band and the measurement noise in which the outlier values have been removed.

Using~\eqref{eq:obs_mdl_2} and the properties of the vectorization operator, we can write this model equivalently as
\begin{align}
    \widetilde{\by}_{k,\ell}^{m}
    &= \big[(\bc_{\ell}^{m})^\top \otimes \bD_k^{m} \big]  \vect\big({\calH}_{\ell}^{m}(\bS_k)\big) + \widetilde{\br}_{k,\ell}^{m}
    \nonumber \\
    &= \big[(\bc_{\ell}^{m})^\top \otimes \bD_k^{m} \big]  \bH_{\ell}^{m}\bs_k + \widetilde{\br}_{k,\ell}^{m}
\end{align}
where $\otimes$ denotes the Kronecker product and $\bs_k=\vect(\bS_k)$ and $\bH_{\ell}^{m}$ is a matrix form representation of the operator ${\calH}_{\ell}^{m}$, such that $\vect({\calH}_{\ell}^{m}(\bS_k))=\bH_{\ell}^{m}\bs_k$.

We can now represent all bands from each modality in the form of a single vector $\widetilde{\by}_{k}^{m}\in{\amsmathbb{R}}^{\widetilde{N}_mL_m}$ as
\begin{align}
    \widetilde{\by}_{k}^{m} = 
    \underbrace{\begin{pmatrix}
    \big[(\bc_{1}^{m})^\top \otimes \bD_k^{m} \big] \bH_{1}^{m} \\
    \vdots\\
    \big[(\bc_{L_m}^{m})^\top \otimes \bD_k^{m} \big] \bH_{L_m}^{m}
    \end{pmatrix}}_{\widetilde{\bH}_k^{m}} \bs_k + \widetilde{\br}_{k}^{m} \,,
    \label{eq:meas_mdl_4}
\end{align}
where $\widetilde{\br}_{k}^{m}\sim{\calN}(\cb{0},\widetilde{\bR}_k^m)$, and
\begin{align}
    \widetilde{\by}_{k}^{m} &= \operatorname{col}\big\{\widetilde{\by}_{k,1}^{m},\ldots,\widetilde{\by}_{k,L_m}^{m}\big\} \,,
    \\
    \widetilde{\br}_{k}^{m} &= \operatorname{col}\big\{\widetilde{\br}_{k,1}^{m},\ldots,\widetilde{\br}_{k,L_m}^{m}\big\} \,,
    \\
    \widetilde{\bR}_k^m &= \operatorname{blkdiag}\big\{\bD_k^{m}\bR_{1}^m(\bD_k^{m})^\top, \ldots, \bD_k^{m}\bR_{L_m}^m(\bD_k^{m})^\top\big\}
\end{align}
Note that at most time instants $k$, one or more of the modalities $m\in\Omega$ is not observed. In this case, we set the matrix $\bD_k^{m}$ as an empty (zero-dimensional) matrix, which simplifies the problem and avoids introducing additional notation.

\subsection{Dynamical evolution model}
Defining reasonable dynamical models for image fusion requires detailed knowledge regarding the scene evolution over time, which is often unattainable. 
In this contribution, we aim at a complete data driven strategy assuming very little knowledge regarding the scene evolution except for past data coming from the imaging modalities being used. To match such lack of prior knowledge we consider a simple random-walk process to model the latent state dynamics as:
\begin{align}
    \bs_{k+1} = \bF_k\bs_k + \bq_k \,,
    \label{eq:state_evol_mdl_1}
\end{align}
where $\bF_k\in{\amsmathbb{R}}^{L_H N_H}$ is the state transition matrix, which is assumed to satisfy $\|\bF_k\|_2\leq1$, and $\bq_k\sim{\calN}(\cb{0},\bQ_k)$ with $\bQ_k\in{\amsmathbb{R}}^{L_H N_H\times L_H N_H}$ being the {process noise} covariance matrix. {In this paper, $\bF_k$ is set to be {the identity matrix}.}
Note that the above model plays a crucial role in the estimation results, as it describes both the distribution of the changes occurring in the image at time $k$, as well as the marginal distribution of the states. 
This means that more sophisticated dynamics can be introduced in the problem through the appropriate design of the innovation covariance matrix $\bQ_k$.
%
{Although expectation maximization (EM)~\cite{sarkka2013bayesianBook,borsoi2020multitemporalUKalmanEM} or maximum likelihood estimation (MLE) based methods~\cite{mehra1972approaches}} can be used to estimate $\bQ_k$ in time invariant models {(i.e., when $\bQ_j=\bQ_k$ for all $j,k$)}, the problem becomes extremely ill-posed in the time-varying setting.
Another issue {is related} to the computational complexity of {MLE or} EM-based strategies {which, e.g., require} the solution of the Kalman filter and smoother systems {to be computed} multiple times, becoming unfeasible when dealing with large images. {Moreover, such strategies do not directly leverage historical data.}
%
For these reasons, we propose an alternative route to estimate $\bQ_k$.

\subsubsection{Data-driven approach for estimating $\bQ_k$}\label{sec:Qest}

We consider $\bQ_k(\cp{D}_k)$ as a function of the set  $\cp{D}_k=\{\tilde{\by}_\ell^{m\in\Omega_H}\}_{\ell<k}$ of past high resolution images. The set $\cp{D}_k$ represents historical data and images currently being fused up the the time step $k$.
Although many strategies could be leveraged to find suitable past time windows to account for more relevant covariance estimation and consider full covariance matrices, in this preliminary work we choose a simple route to validate this type of approach. 
For this, let ${\by}^{m\in\Omega_H}_{k-\tau}$ be the the most recently observed high resolution image\footnote{That is, $\tau\in{\amsmathbb{Z}_+}$ is the smallest integer such that a high resolution image was observed at time instant $k-\tau$.}. We compute $\bQ_k$ by finding in our historical data the most similar image to ${\by}^{m\in\Omega_H}_{k-\tau}$ and then computing the pixelwise variance across the next $d\in\bbZ_+$ images in our historical data. That is, we compute $\bQ_k$ executing the following three steps for every time step $k$:
\begin{enumerate}
    \item Identify the most similar state over $\cp{D}_k$, that is, the image that is most similar, according to a metric $\cal{L}$
    \begin{equation}
        \ell^* = \mathop{\arg\min}_{\ell \in {\cp{I}}_{\cp{D}_k}} \,\, {\cal{L}}\big({\by}^{m\in\Omega_H}_{k-\tau}, [{\cal{D}}_k]_\ell\big) \,,
    \end{equation}
    with $[{\cal{D}}_k]_\ell$ being the $\ell$-th image in the historical set ${\cal{D}}_k$, and ${\cp{I}}_{\cp{D}_k}\subseteq{\amsmathbb{Z}}$ is the set containing the time index of each image in ${\cp{D}_k}$.
    
    \item select a time window $[{{\calD}_k}]_{\ell^*:\ell^*+d}$.
    \item compute the diagonal innovation matrix, i.e.,  $\bQ_k=\operatorname{diag}\{q_{k,1}^2, \ldots, q^2_{k,L_HN_H}\}$, as
    \begin{equation}
    q^2_{k,j} = \max\big(\operatorname{var}([{\cal{D}}_k]_{\ell^*:\ell^* + d}^{(j)}), \varepsilon^2\big) \,,
\end{equation}
\end{enumerate}
where $[{\cal{D}}_k]_{\ell^*:\ell^*+d}^{(j)} = [\tilde{y}_{\ell^*,j}^{m\in\Omega_H},\ldots, \tilde{y}_{\ell^*+d,j}^{m\in\Omega_H}]$, and $\varepsilon>0$ is a small scalar allowing for changes on the scene that were unseen on the historical data window $[{\cal{D}}_k]_{\ell^*:\ell^*+d}$. 
As similarity metric we used the cosine similarity $\cal{L}(\by, \bz) = \cos(\by,\bz)$.

\section{Multimodal image fusion using the Kalman filter}
\label{sec:filter}

Considering models \eqref{eq:meas_mdl_4} and \eqref{eq:state_evol_mdl_1}, the online multimodal image fusion problem can be formulated as the problem of computing the posterior distribution of the high resolution image given all previous measurements available, i.e.,
\begin{align}
    p\big(\bs_k\big|\{\widetilde{\by}_{1:k}^{m}\}_{m\in\Omega}\big) = {\calN}\big(\bs_{k|k},\bP_{k|k}\big)\,.
    \label{eq:posterior_1}
\end{align}
%
Due to the choice of a linear Gaussian model, this distribution is also Gaussian. Moreover, its mean vector $\bs_{k|k}$ and covariance matrix $\bP_{k|k}$ can be computed recursively using the standard Kalman filter with a prediction and update steps~\cite{sarkka2013bayesianBook}.


More precisely, the prediction step of the Kalman filter computes the first and second order moments of $p\big(\bs_k\big|\{\widetilde{\by}_{1:k-1}^{m}\}_{m\in\Omega}\big)$ as:
\begin{align}
    \bs_{k|k-1} &= \bF_{k-1}\bs_{k-1|k-1}\\
    \bP_{k|k-1} &= \bF_{k-1}\bP_{k-1|k-1}\bF_{k-1}^\top + \bQ_{k-1}
\end{align}
The update step computes then computes of~\eqref{eq:posterior_1}. Note that the update step can be simplified and implemented separately for each data modality by using the Markov property of the model and the independence between noise vectors of different modelities:
\begin{align}
    p\big(\bs_k\big|&\{\widetilde{\by}_{1:k}^{m}\}_{m\in\Omega}\big)
    \nonumber \\
    & \propto
    p\big(\{\widetilde{\by}_k^{m}\}_{m\in\Omega}\big|\bs_k\big) 
    \nonumber 
    p\big(\bs_k\big|\{\widetilde{\by}_{1:k-1}^{m}\}_{m\in\Omega}\big)
    \\
    & = p\big(\bs_k\big|\{\widetilde{\by}_{1:k-1}^{u}\}_{u\in\Omega}\big) \prod_{m\in\Omega} p\big(\widetilde{\by}_k^{m}\big|\bs_k\big) \,.
    \label{eq:multi_upd_1}
\end{align}
By computing the first product in the right hand side as:
\begin{align}
    & p\big(\bs_k\big|\{\widetilde{\by}_{1:k-1}^{u}\}_{u\in\Omega}\big) p\big(\widetilde{\by}_k^{m}\big|\bs_k\big) 
    \nonumber\\
    & \hspace{7ex} \propto p\big(\bs_k\big|\{\widetilde{\by}_{1:k-1}^{u}\}_{u\in\Omega},\widetilde{\by}_k^{m}\big) \,,
    \label{eq:multi_upd_2}
\end{align}
which is an update step of the Kalman filter with image modality $m$ to yield a new posterior in the r.h.s. of~\eqref{eq:multi_upd_2}. This can be computed as:
\begin{align}
    \bv_k^m &= \widetilde{\by}_{k}^{m} - \widetilde{\bH}_k^{m} \bs_{k|k-1}\\
    \bT_k^m &= \widetilde{\bH}_k^{m}\bP_{k|k-1}\big(\widetilde{\bH}_k^{m}\big)^\top + \widetilde{\bR}_k^m\\
    \bK_k^m &=  \bP_{k|k-1} \big(\widetilde{\bH}_k^{m}\big)^\top \big(\bT_k^m\big)^{-1} \\
    \bs_{k|k} &= \bs_{k|k-1} + \bK_k^m \bv_k^m\\
    \bP_{k|k} &=  \bP_{k|k-1} - \bK_k^m\bT_k^m \big(\bK_k^m\big)^\top
\end{align}
for $m\in\Omega$. By proceeding with the computation of the product in the r.h.s. of~\eqref{eq:multi_upd_1} recursively, the Kalman update can then be performed separately for each of the modalities observed at time instant~$k$. Note that after the first modality is processed, the update equations above are used again for the subsequent modalities by setting $\bs_{k+1|k}$ and $\bP_{k+1|k}$ as equal to the posterior estimates from the previously processed modality.

\subsection{The Linear Smoother}

Given a window of $K$ image samples, the Bayesian smoothing problem consists of computing the posterior distribution of the high resolution image given all available measurements available, i.e.,
\begin{align}
    p\big(\bs_k\big|\{\widetilde{\by}_{1:K}^{m}\}_{m\in\Omega}\big) = {\calN}\big(\bs_{k|K},\bP_{k|K}\big) \,,
    \label{eq:posterior_2}
\end{align}
which is also a Gaussian.
Just like in the filtering problem, the linear and Gaussian model allows this solution to be computed efficiently using the Rauch-Tung-Striebel (RTS) smoothing equations~\cite{sarkka2013bayesianBook}, which consist of a forward pass of the Kalman filter (as described before), followed by a backwards recursion that updates the previously computed mean and covariances matrices of the state with information from future time instants.

We note that the smoothing can also be performed efficiently for the case when multiple image modalities are available. Let us consider the Bayesian smoothing equations as defined in~\cite{kitagawa1987smoother,sarkka2013bayesianBook}, which is performed in two steps. Starting from the Kalman state estimate at time $K$, given by $p\big(\bs_{K}\big|\{\widetilde{\by}_{1:K}^{m}\}_{m\in\Omega}\big)$, the smoothing distribution is computed recursively for $k=k-1,\ldots,1$, according to the following relation:
\begin{align}
    p\big(\bs_{k}\big|\{&\widetilde{\by}_{1:K}^{m}\}_{m\in\Omega}\big) = p\big(\bs_k\big|\{\widetilde{\by}_{1:k}^{m}\}_{m\in\Omega}\big)
    \nonumber\\
    & \times \int \frac{p(\bs_{k+1}|\bs_k)p\big(\bs_{k+1}\big|\{\widetilde{\by}_{1:K}^{m}\}_{m\in\Omega}\big)}{p\big(\bs_{k+1}\big|\{\widetilde{\by}_{1:k}^{m}\}_{m\in\Omega}\big)} d\bs_{k+1} \,,
\end{align}
where $p\big(\bs_k\big|\{\widetilde{\by}_{1:k}^{m}\}_{m\in\Omega}\big)={\calN}(\bs_{k|k},\bP_{k|k})$ is the Kalman estimate of the state PDF at time $k$, $p(\bs_{k+1}|\bs_k))$ is the state transition PDF, computed according to~\eqref{eq:state_evol_mdl_1}, $p\big(\bs_{k+1}\big|\{\widetilde{\by}_{1:K}^{m}\}_{m\in\Omega}\big)={\calN}(\bs_{k+1|K},\bP_{k+1|K})$ is the smoothing distribution obtained at the previous iteration, and $p\big(\bs_{k+1}\big|\{\widetilde{\by}_{1:k}^{m}\}_{m\in\Omega}\big)$ is the predictive state distribution, which is computed exactly as in the prediction step of the Kalman filter.

In the linear and Gaussian case this translates into the following closed form solution~\cite{sarkka2013bayesianBook}, with 
\begin{align}
    \bs_{k+1|k} &= \bF_k\bs_{k|k} \\
    \bP_{k+1|k} &= \bF_k\bP_{k|k}\bF_k^\top + \bQ_k
\end{align}
being used to compute the predictive state distribution, and
\begin{align}
    \bG_k &= \bP_{k|k}\bF_k^\top \bP_{k+1|k}^{-1}
    \\
    \bs_{k|K} &= \bs_{k|k} + \bG_k(\bs_{k+1|K} - \bs_{k+1|k})
    \\
    \bP_{k|K} &= \bP_{k} + \bG_k(\bP_{k+1|K} - \bP_{k+1|k}) \bG_k^\top
\end{align}
to update the covariances. It should be noted that the mean and covariance $\bs_{k|k}$ and $\bP_{k|k}$ used in the Smoothing equations are the final result obtained from the Kalman update after processing all image modalities that were available at instant~$k$.

Thus, while in the Kalman filtering the update equations must be computed sequentially at each time step w.r.t. the different image modalities, smoothing only needs only the final state estimates at each instant, no matter how many modalities are present.


\subsection{Reducing the computational complexity}\label{sec:complex}

A problem with the Kalman filter is the need to compute and store the state covariance matrix, $\bP_{k|k}$. This incurs in storage and operations complexity in the order of $O(N_H^2 L_H^2)$ and $O(N_H^3 L_H^3)$, respectively. This can make the method intractable for images with a large number of pixels. Thus, to reduce the complexity of the filter and of the smoother, we consider a patch-based strategy for processing large images. More precisely, we divide the high and the low resolution images into different corresponding patches containing a smaller number of pixels. Assuming the pixels in different patches to be statistically independent, we can consider the dynamical model in Section~\ref{sec:model} individually for each patch and thus apply the filtering and smoothing framework with a reduced computational complexity.




\section{Experiments}
\label{sec:experiments}

In this section, we use the proposed methodology to fuse Landsat and MODIS image over time, which we denote by KFQ and SMQ for the filter and for the smoother, respectively. Although in our experiments we consider only two modalities the proposed methodology admits multiple different modalities provided that enough computational power is available.   
As benchmark, we compare the performance of the Kalman Filter (KF), Smoother (SM) methods with constant $\bQ_k = \xi\bI$ and with data-driven innovation matrix, see, Section~\ref{sec:Qest}, KFQ and SMQ to that of the \emph{Enhanced Spatial and Temporal Adaptive ReFlectancefusion Model} (ESTARFM) algorithm~\cite{zhu2010fusion_ESTARFM}.
In the following, we describe the data and simulation setup, followed by the results and the discussions.


\subsection{Study region}

\begin{figure}[h!]
  \centering
\includegraphics[width=1\linewidth]{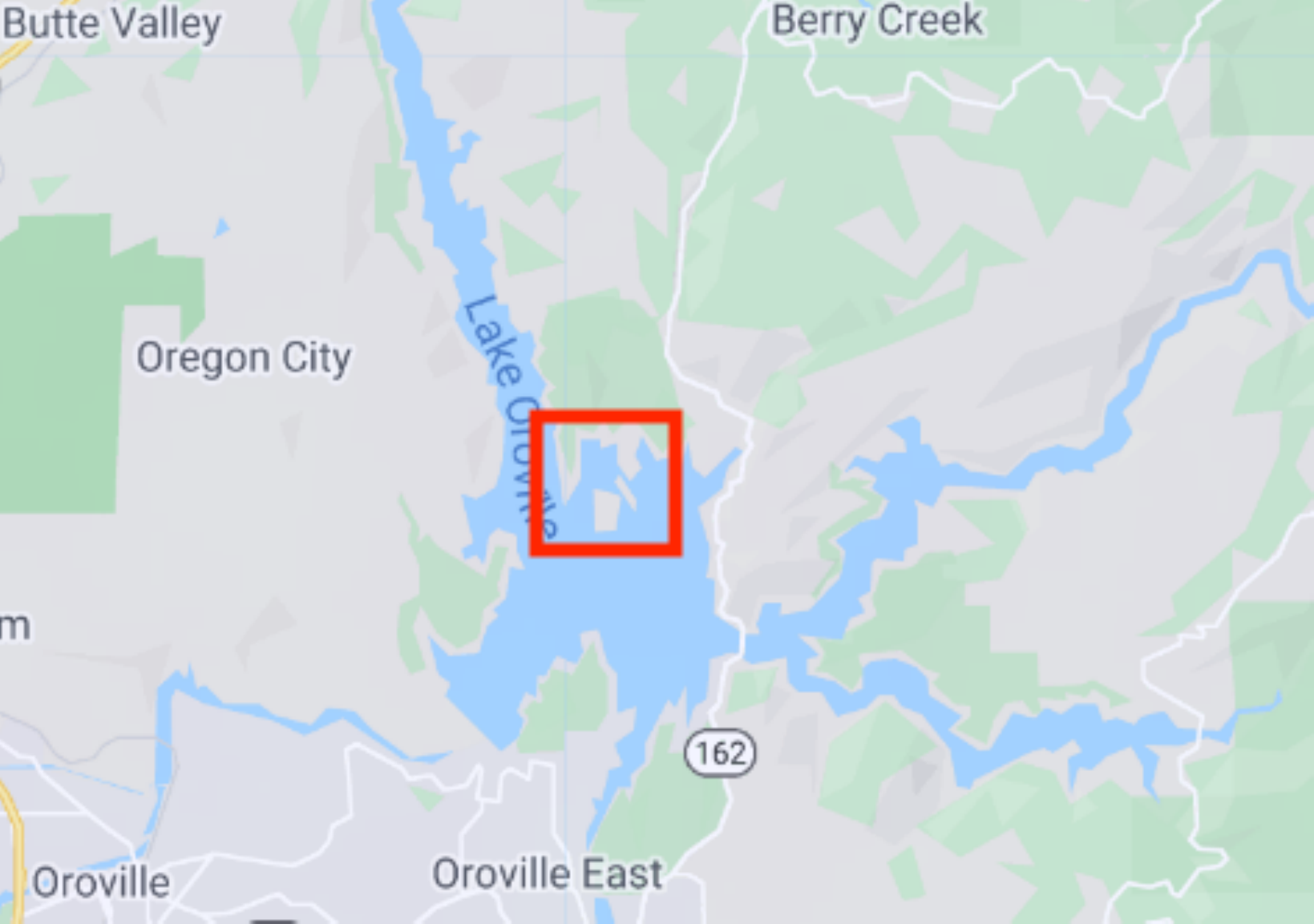}
  \caption{Oroville dam. The red box delimits the specific the study area used in our experiments.}
  \label{fig:site_location}
\end{figure}

\begin{figure}[h!]
  \centering
  \includegraphics[width=1\linewidth]{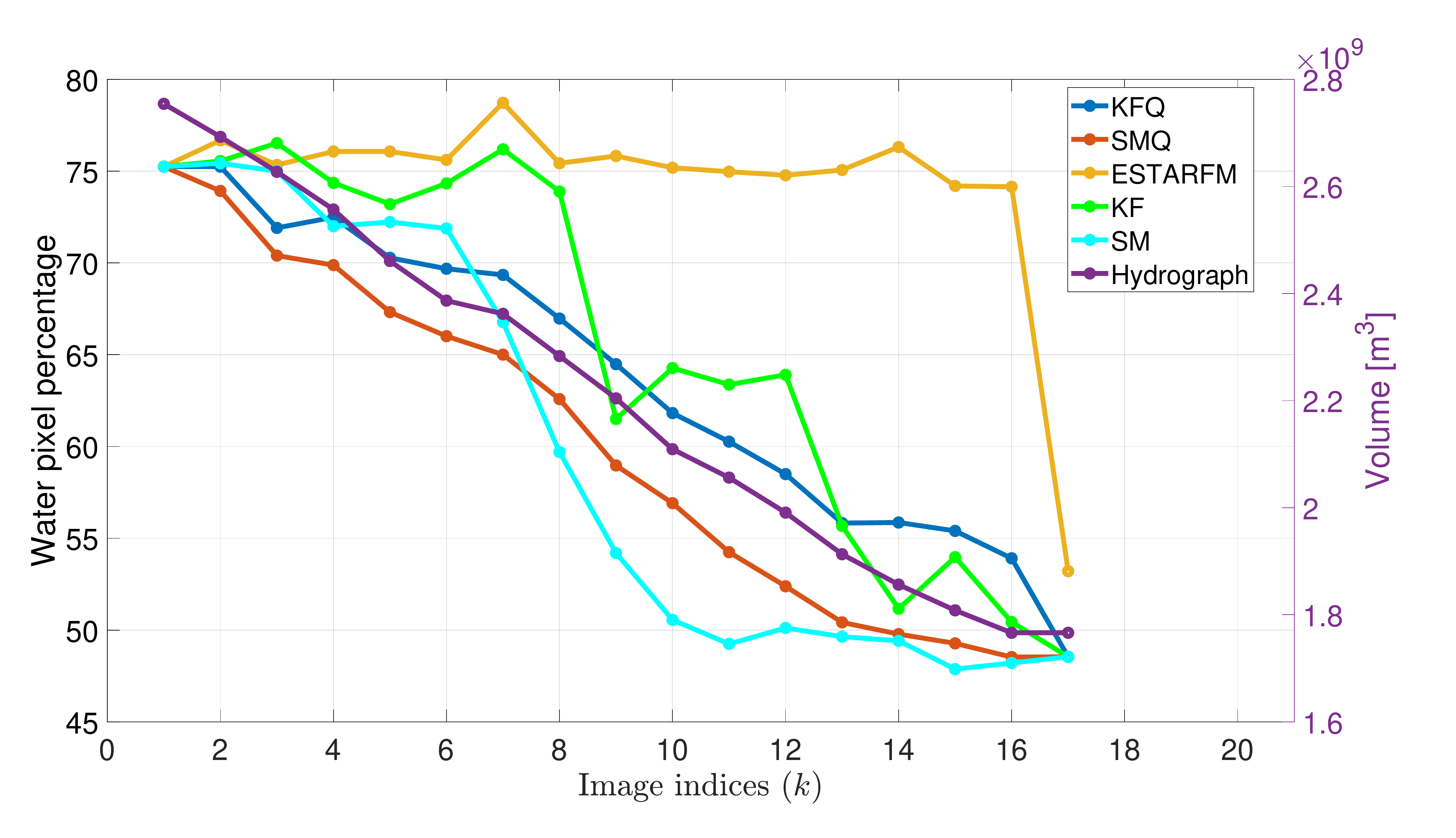}
  \vspace{-0.7cm}
  \caption{Percentage of water pixels in the estimated images over image index (time) and the reservoir volume in $m^3$ (hydrograph). Classification of water was done by performing clustering the estimated bands for each method and time index. High resolution Landsat images were observed at index $k\in\{1, 17\}$. We observe that KFQ and SMQ match the hydrograph curve much more closely than the other algorithms.}
  \label{fig:hydrographest}
\end{figure}

For the experiments, we consider the Oroville dam (Figure~\ref{fig:site_location}), located on the Feather River, in the Sierra Nevada Foothills (38° 35.3' North and 122° 27.8' W) is the tallest dam in USA and is major water storage facility in California State Water Project. The reservoir has a maximum storage capacity of $1.54\times10^{11}$ ft$^{3}$ or $4.36\times10^{9}$ m$^{3}$, which fills during heavy rains or large spring snow melts and water is carefully released to prevent flooding in downstream areas, mainly to prevent large flooding in Butte County and area along the Feather River. The reservoir water storage change in between 07/03 and 09/21 of 2018 is as shown as the hydrograph curve in Figure~\ref{fig:hydrographest}. Another unique characteristic is that it has three power plants at this reservoir. The water released downstream is used to maintain the Feather and Sacramento Rivers and the San Francisco-San Joaquin delta. Lake Oroville is at an elevation of 935 feet (285 meters) above sea level. 
We focus at a particular area of the Oroville dam delimited by the red box in Figure~\ref{fig:site_location}.


\begin{figure*}[h!]
  \centering
  \begin{subfigure}[b]{1\textwidth}
  \includegraphics[width=1\linewidth]{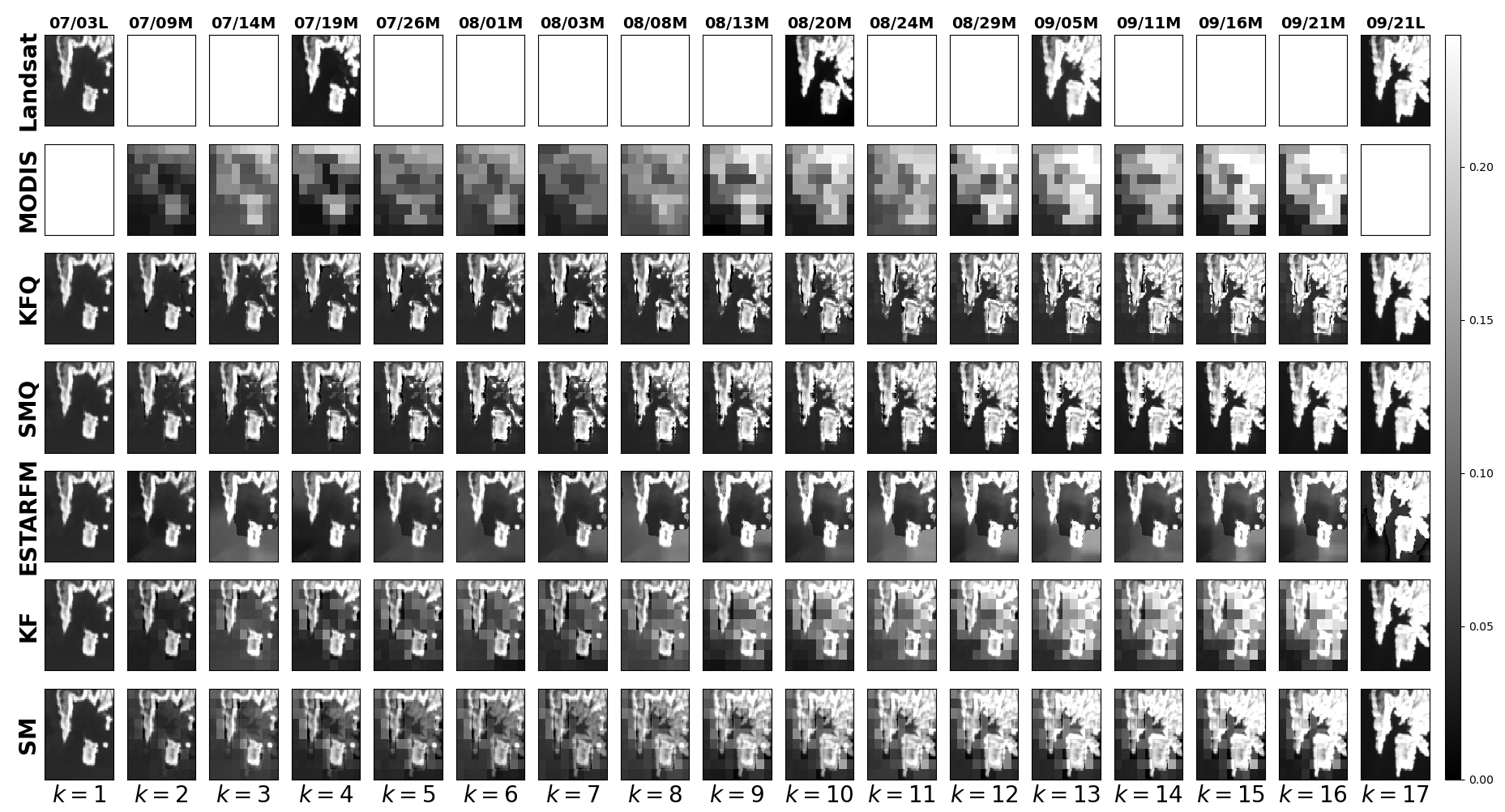}
  \caption{Fused images in band 1 (MODIS) and band 4 (LandSat)}
  \label{fig:Reconstruction1}
\end{subfigure}
\\
\begin{subfigure}[b]{1\textwidth}
  \centering
  \includegraphics[width=1\linewidth]{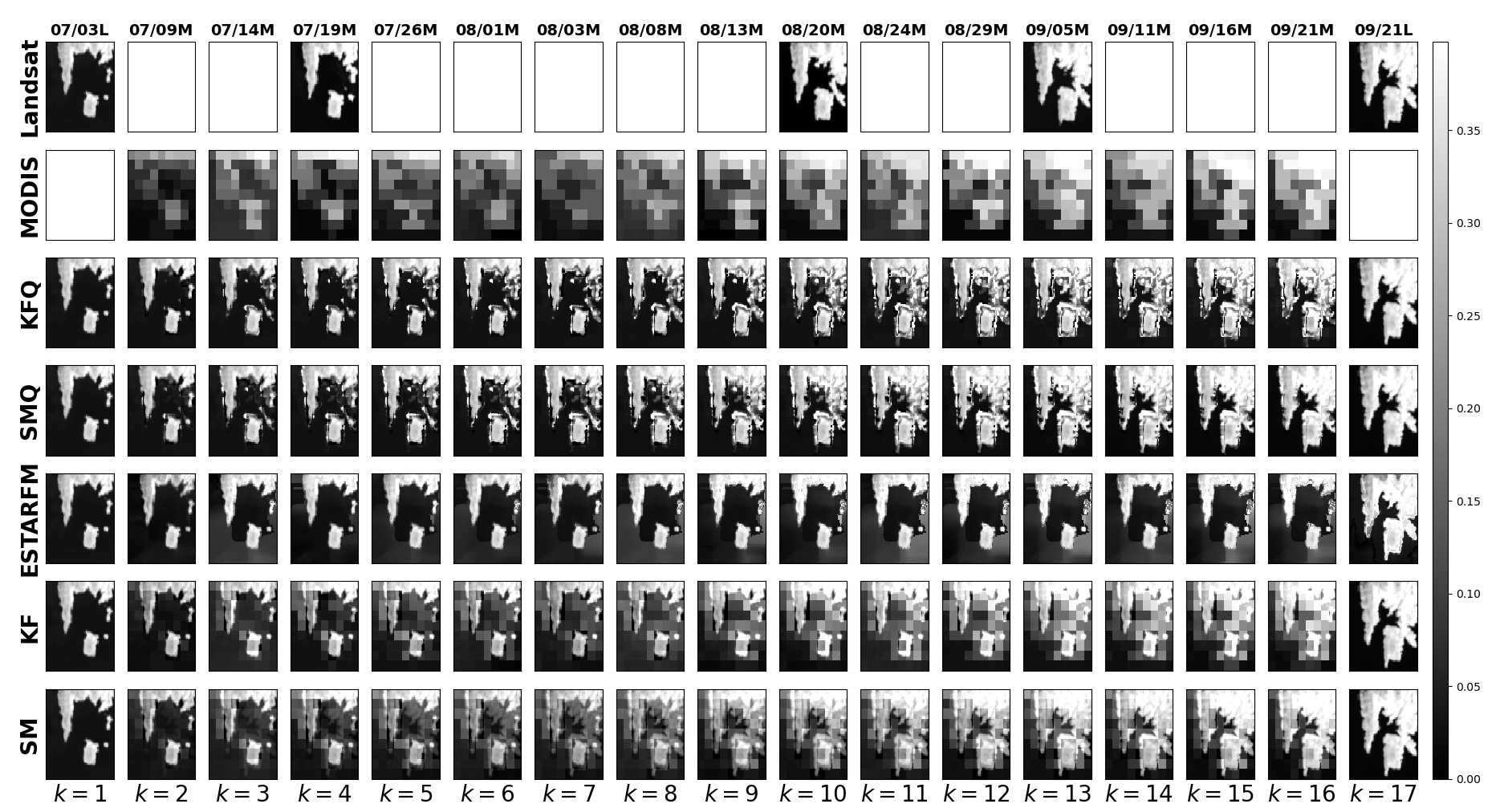}
  \caption{Fused images in band 2 (MODIS) and band 5 (LandSat)}
  \label{fig:Reconstruction2}
  \end{subfigure}
  \caption{Fused bands from MODIS and Landsat using different strategies over time. The first two rows of each subfigure depict MODIS and Landsat bands acquired at dates displayed on top labels. At each time index estimation with KF, SM, KFQ, SMQ and ESTARFM are presented. Some Landsat images were omitted from the estimation process and used solely as ground truth. Images used at each update step are indicated on top labels where ``M'' stands for MODIS and ``L'' for Landsat.}\label{fig:Reconstruction}
\end{figure*}

\begin{figure*}[h]
\begin{subfigure}[b]{1\textwidth}
  \centering
  \includegraphics[width=.8\linewidth]{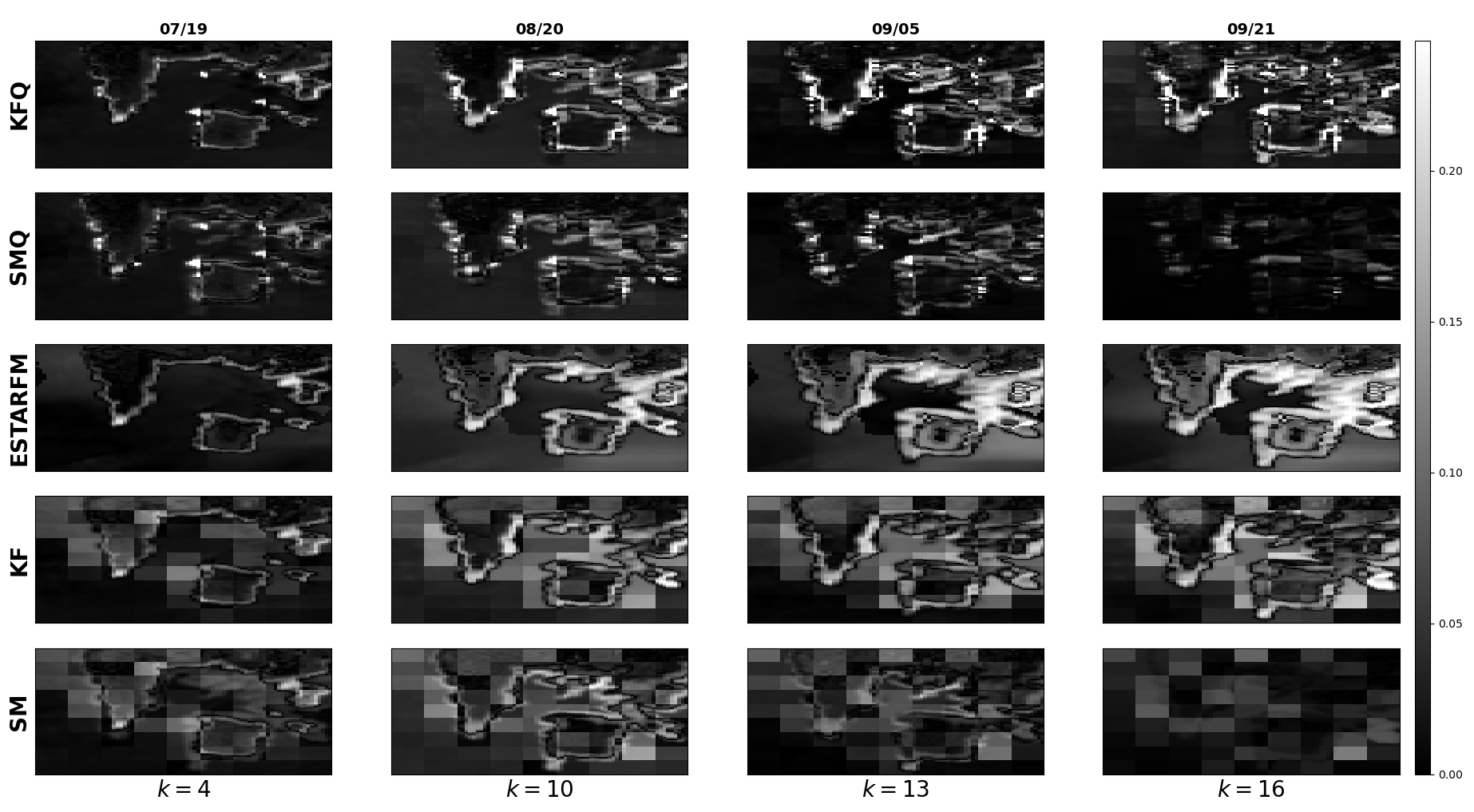}
  \caption{Error map of images in band 1 (MODIS) and band 4 (Landsat).}
  \label{fig:errormap1}
  \end{subfigure}
%
\begin{subfigure}[b]{1\textwidth}
  \centering
  \includegraphics[width=.8\linewidth]{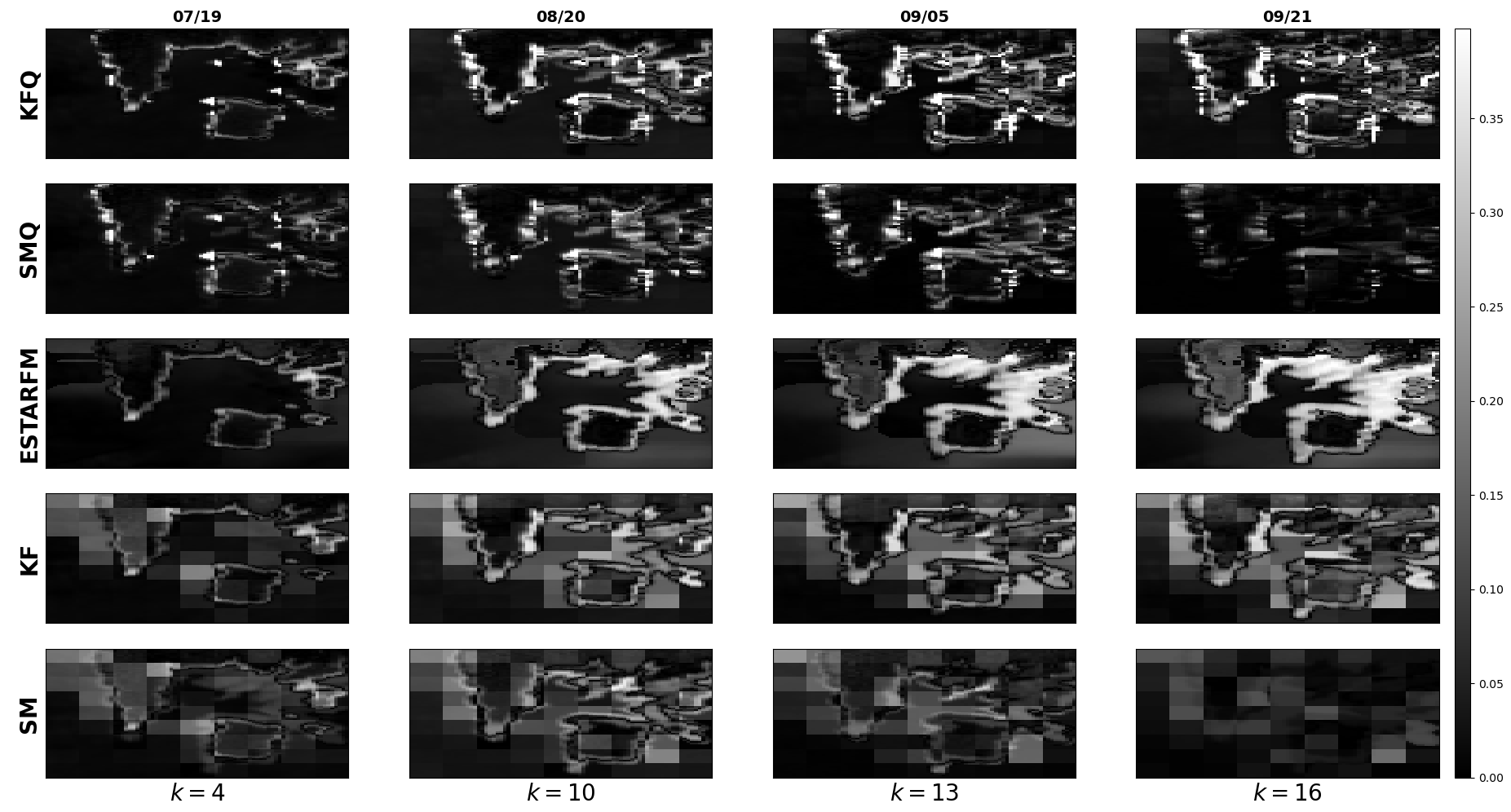}
  \caption{Error map of images in band 2 (MODIS) and band 5 (Landsat).}
  \label{fig:errormap2}
  \end{subfigure}
  \caption{Absolute error maps for red (Figure~\ref{fig:errormap1}) and NIR (Figure~\ref{fig:errormap2}) bands.} \label{fig:ErrorMap}
\end{figure*}

\begin{figure*}[h]
  \centering
  \includegraphics[width=.8\linewidth]{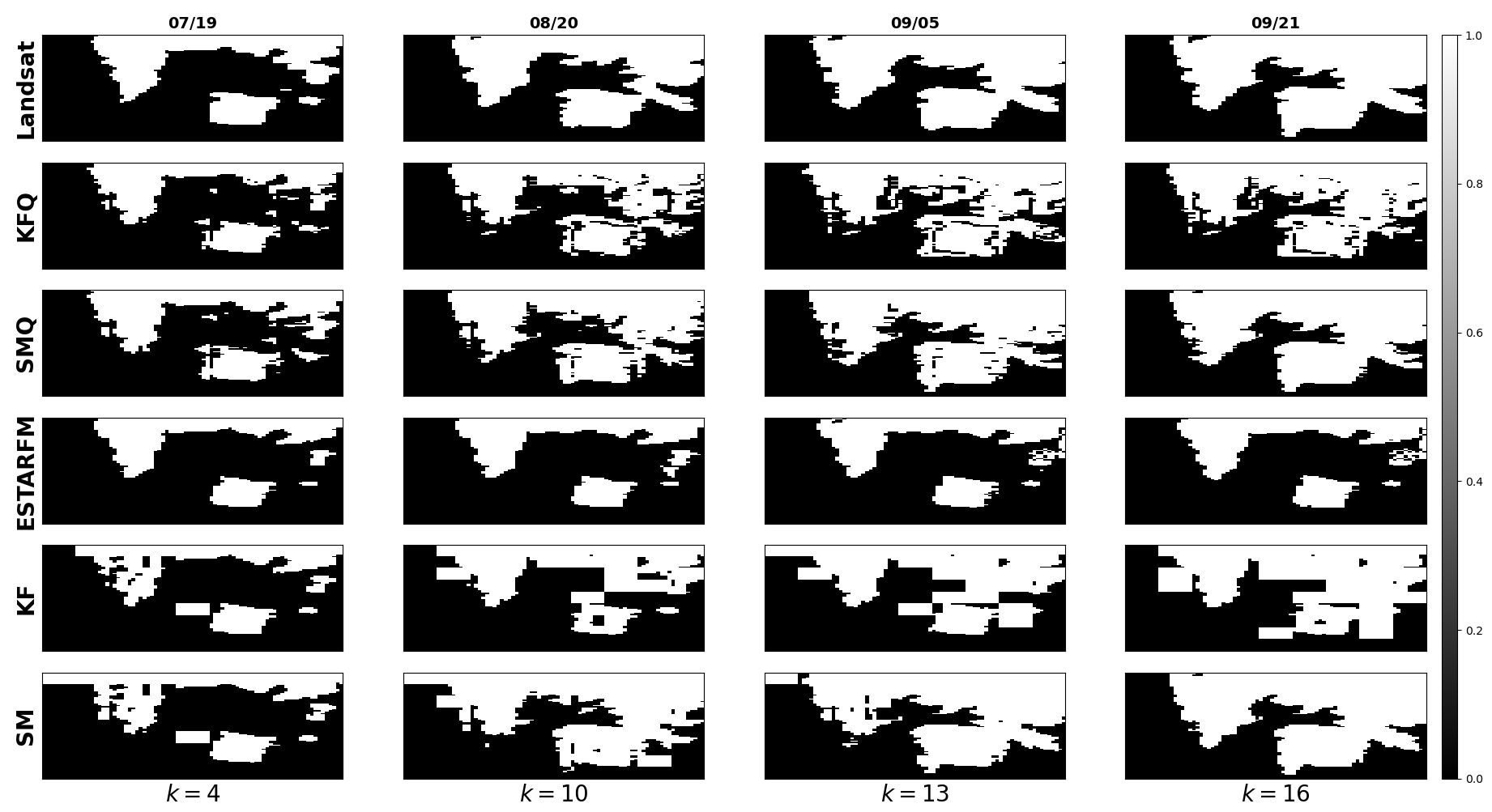}
  \caption{Water map of images based on K-means clustering strategy where 1 indicates land and 0 indicates water pixels. Unused Landsat classification maps establish the ground-truth (first row). KFQ and SMQ present classfication maps that are semantically better than the competing methods. Although SM presents classification maps containing various square patches, they present the second best results in absolute terms, see, Table~\ref{tab:miscount}, followed by KFQ, KF and ESTARFM.}
  \label{fig:watermap}
\end{figure*}
\begin{table} [!ht]
\scriptsize
\centering
\caption{Percentage of mis-classified pixels.}
\medskip
\vspace{-0.45cm}
\renewcommand{\arraystretch}{1.2}
\resizebox{\linewidth}{!}{%
\begin{tabular}{c||c|c|c|c|c}
\hline
Method & KFQ & SMQ & ESTARFM & KF & SM \\
\hline\hline
{Image (07/19)} & 6.1423 & \textbf{5.4260}  & 5.4870 &8.0171 &9.4041 \\
\hline
{Image (08/20)} & 10.0899 & \textbf{7.0416}  & 18.3051 &16.4761 &13.6717\\
\hline
{Image (09/05)} & 12.0408 & \textbf{7.5446} & 22.7404 &18.1375 &8.1390  \\
\hline
{Image (09/21)} & 9.8461  & \textbf{1.7528}  & 26.3374 &18.9605 &2.3777 \\
\hline\hline
{Average} & 9.5298 & \textbf{5.4412} & 18.2175 &15.3978 &8.3981 \\
\hline
\end{tabular}
}
\label{tab:miscount}
\end{table}

\begin{table} [!ht]
\scriptsize
\centering
\caption{Average Normalized Rooted Mean Square Error (NRMSE) between the true and estimated Landsat images.}
\renewcommand{\arraystretch}{1.2}
\resizebox{\linewidth}{!}{%
\begin{tabular}{c||c|c|c|c|c}
\hline
 Method & {KFQ} & {SMQ} & ESTARFM & {KF} & {SM} \\
\hline	\hline		
{Image (07/19)} & 0.3672 & 0.3815  & \textbf{0.2905} &0.4157 &0.4269 \\
\hline
{Image (08/20)} & 0.5071 & \textbf{0.4049}  & 0.5517 &0.5355 &0.4611\\
\hline
{Image (09/05)} & 0.5016 & 0.3426 & 0.5775 &0.4794 &\textbf{0.3371}  \\
\hline
{Image (09/21)} & 0.4940  & \textbf{0.1466}  & 0.6125 &0.4914 &0.2200 \\
\hline\hline
Average  &0.4675 &\textbf{0.3189} &0.5081 & 0.4805 &0.3613\\
\hline
\end{tabular}
}
\label{tab:NRMSE}
\end{table}

\subsection{Remote Sensed data}

For our simulations we collected MODIS and Landsat data acquired from the region marked with a red square on Figure~\ref{fig:site_location}, and on a interval ranging from $2018/07/03$ to $2018/09/21$. This interval was selected since the hydrograph analysis indicates high variation in the water level of the reservoir, see, the hydrograph curve in Figure~\ref{fig:hydrographest}. Such variation in the water levels result in large changes in the acquired images, exposing flooded areas. In this experiment we will focus on the red and near-infrared (NIR) bands since they are often used to distinguish water from other landcover elements in the image~\cite{gao1996ndwi}. We also collected $5$ Landsat data from $2017/08/01$ to $2017/12/07$ to serve as a past historical dataset~$\cp{D}_k$.

\interfootnotelinepenalty=10000

The study region marked in Figure~\ref{fig:site_location} corresponds to Landsat and MODIS images with $81\times 81$ and $9\times 9$ pixels, respectively\footnote{The Landsat images were also upsampled to a spatial resolution of 27.77 meters to make its resolution exactly 9 times that of MODIS.}. After filtering for heavy cloud cover during the designated time periods, a set of 6 landsat and 16 MODIS images were obtained. We used the first MODIS and Landsat images for initialization of all methods leading to 5 and 15 images used in the remaining fusion process. 


From the set of 5 Landsat images that were available for testing, three of them were set aside and not processed by any of the the algorithms. These images were acquired at dates 07/19, 08/20 and 09/05, when MODIS observations were also available, and will be used in the form of a reference for the evaluation of the algorithms' capability of estimating the high resolution images at these dates solely from the low resolution MODIS measurements,
since the noise in the Landsat images is comparatively low.

\subsection{Algorithm setup}

We initialized the proposed Kalman filter and smoother (KFQ and SMQ) using a high resolution Landsat observation as the state, i.e., $\bs_{0|0}=\widetilde{\by}_0^{\mathsf{L}}$, and set $\bP_{0|0}=10^{-10}\bI$. The noise covariance matrices were set as $\bR_{\ell}^{\mathsf{L}}=10^{-10}\bI$ and $\bR_{\ell}^{\mathsf{M}}=10^{-4}\bI$, for all $\ell$. The blurring and downsampling matrices were set as $\bH_{\ell}^{\mathsf{L}}=\bI$ for Landsat, while for MODIS $\bH_{\ell}^{\mathsf{M}}$ consisted of a convolution by an uniform $9\times9$ filter, following by decimation by a factor of~$9$, which represents the degradation occurring at the sensor (see, e.g.,~\cite{huang2013spatiotemporalFusionBayes}). We also set $\bF_k=\bI$ for all $k$. The vectors $\bc_{\ell}^m$ contained a positive gain in the $\ell$-th position which compensated for scaling differences between Landsat and MODIS sensors, and zeros elsewhere.

The matrices $\bD_k^{\mathsf{M}}$ were constructed based on the quality codes (i.e., the QA bits) released by MODIS for each image pixel~\cite{Vermote_modis_2015}. 
QA bits provides information regarding pixel quality and cloud cover for all pixels and all bands. In our experiments we dropped any pixel not classified as \emph{corrected product produced at ideal quality} in the QA bits~\cite{Vermote_modis_2015} by adding zeros at corresponding positions in $\bD_k^{\mathsf{M}}$.
Matrices $\bQ_k$ were computed following our data-driven strategy described in Section~\ref{sec:Qest} where $\varepsilon^2=10^{-5}$ and $d=1$.
The setup of the benchmark Kalman filter and smoother (KF and SM) was identical to the one of KFQ and SMQ described above, except for having a fixed $\bQ_k=10^{-2}\bI$.

The ESTARFM algorithm was parametrized as follows \cite{zhu2010fusion_ESTARFM}, $w = 14$ as half of the window size, the number of classed was set to $4$, and the pixels range was set to $[0, 0.5]$.

All algorithms are evaluated using three metrics, which are computed taking as reference the Landsat images, three of which are not observed by the algorithms. The first metric is the normalized root mean squared error (NRMSE), which attempts to measure the estimation accuracy directly:
\begin{align}
    \operatorname{NRMSE}(\bs,\widehat{\bs}) = \sqrt{\frac{\|\bs-\widehat{\bs}\|^2}{\|\bs\|^2}}
\end{align}
where $\bs$ and $\widehat{\bs}$ denote the true and the estimated images, respectively. The two remaining metrics are related to downstream tasks of water classification and water level monitoring, which are performed on the reconstructed image sequence. 

We evaluate the direct benefit of the analyzed strategies in classifying water pixels from the estimated images. To classify water pixels we resorted to a KNN classifier whose centroids of water and non-water 2-band pixels were computed using K-Means algorithm. 
Finally, we evaluate the performance of the algorithms for hydrograph estimation by plotting the proportion of pixels in the image classified as water over time against the true hydrograph for the period, for all algorithms.





\subsection{Results}

As discussed, we fused the red and NIR reflectance bands of MODIS and Landsat for the selected study region. In Figure~\ref{fig:Reconstruction}, we show the fused red (Figure~\ref{fig:Reconstruction1}) and NIR (Figure~\ref{fig:Reconstruction2}) reflectances as well as the acquired red and NIR reflectance values from MODIS and Landsat. Acquisition dates are displayed in the top labels at each column with a character, $M$ for MODIS and $L$ for Landsat indicating the image used in the fusion algorithms. We recall that only the first and last Landsat images were used in the fusion process, keeping the remaining three images as ground-truth for evaluation purposes. 
Analyzing the results we can see that the images estimated by the proposed KFQ and SMQ methods produce better visual similarity with the Landsat (ground-truth) images for both bands. For instance, the increase in the island and the expansion of other land parts are clearly visible for the proposed methods. In contrast, analyzing ESTARFM results we note that land parts remain mainly constant through time until a new Landsat image is observed. Although lighter areas on the water portions can be noticed, specially for $k>8$, its distribution does not resemble the ground-truth. This is expected since ESTARFM is not designed to acknowledge prior information or historical data. 
When analyzing KF and SM methods, where the innovation matrix was kept constant and independent of past data, we observe that the MODIS contributions appears in patches since the lack of prior information results in changing all the high resolution states within each low resolution pixel range. Nevertheless the KF and SM results seem to recover better land proportion than the ESTARFM.

The results presented above are corroborated by the absolute error maps displayed in Figure~\ref{fig:ErrorMap}, and NRMSE results shown in Table~\ref{tab:NRMSE} for dates in which ground-truth is available.
Analyzing Figure~\ref{fig:ErrorMap} we highlight that SMQ clearly presents the smallest error (i.e., overall darker pixels) for both bands and all dates. KFQ also present low absolute error except for contour regions. SM is the second overall darker image, followed by KF and ESTARFM with exception of the results on 07/19 (first column) where ESTARFM shows low reconstruction error. Similar conclusions can be achieved by analyzing Table~\ref{tab:NRMSE}. 

Figure~\ref{fig:watermap} presents the water maps for the ground-truth (first row) and all studied algorithms obtained using K-means clustering. When comparing the resulting classification maps, KFQ and SMQ presents classification maps that are semantically better than the competing methods. Although SM presents classification maps with square patches they present the second best results in absolute terms, see, Table~\ref{tab:miscount}, followed by KFQ, KF and ESTARFM.

The above water classification results are also corroborated by the mis-classification results presented in Table~\ref{tab:miscount}, where the Kalman filter- and smoother-based methods led to smaller mis-classification rates for all images except the one on 07/19. In terms of average mis-classification rates we can sort the methods from best to worst as SMQ, SM, KFQ, KF, ESTARFM.

Finally, we plotted the percentage of pixels classified as water over the time index $k$ in Figure~\ref{fig:hydrographest}. We can observe that the SMQ and KFQ algorithms lead to curves that are closer to the true hydrograph curve, indicating that the proposed strategies lead to better estimation of water content of the reservoir.




\subsection{Discussion}

The results presented above clearly indicate that including prior information through appropriate innovation matrix design improves the Kalman filter- and smoother-based fusion approaches in both quantitative metrics and semantics of the image evolution over time. For instance, the growth of the island portion over time in regions that \emph{makes sense} leads to more meaningful results that cannot be entirely captured by one standard metric such as NRMSE or mis-classification. Nevertheless, filtering-based strategies clearly outperformed in all absolute metrics the ESTARFM algorithm, which has been the standard algorithm for water mapping with fused low-high resolution satellite data. 
Although our results clearly shows that KFQ and SMQ are better alternatives in all adopted metrics, filtering-based methods demand both computational power and memory specially when dealing with high-dimensional state spaces, see, Section~\ref{sec:complex}, limiting their application to large areas and very high resolution images. Thus, strategies to reduce such computational burden need to be sought and will be addressed in an upcoming work. 



\section{Conclusions}
\label{sec:conclusions}

In this paper, an online Bayesian approach for fusing multi-resolution space-borne multispectral images was proposed.
By formulating the image acquisition process as a linear and Gaussian measurement model, the proposed method leveraged the Kalman filter and smoother to perform image fusion by estimating the latent high resolution image from the different observed modalities. Moreover, a classification-based strategy is also proposed to define an informative time-varying dynamical image model by leveraging historical data, which leads to a better localization of changes occurring in the high-resolution image even in intervals where only coarse resolution observations are available.
Experimental results indicate that the proposed strategy can lead to considerable improvements compared to both the use of noninformative models, and to widely used image fusion algorithms.

\acknowledgments
The authors would like to thank National Geographic, under Grant NGS-86713T-21, for the financial support.

\bibliographystyle{IEEEtran}
\bibliography{references_mtfus}

\begin{thebibliography}{10}
\providecommand{\url}[1]{#1}
\csname url@samestyle\endcsname
\providecommand{\newblock}{\relax}
\providecommand{\bibinfo}[2]{#2}
\providecommand{\BIBentrySTDinterwordspacing}{\spaceskip=0pt\relax}
\providecommand{\BIBentryALTinterwordstretchfactor}{4}
\providecommand{\BIBentryALTinterwordspacing}{\spaceskip=\fontdimen2\font plus
\BIBentryALTinterwordstretchfactor\fontdimen3\font minus
  \fontdimen4\font\relax}
\providecommand{\BIBforeignlanguage}[2]{{%
\expandafter\ifx\csname l@#1\endcsname\relax
\typeout{** WARNING: IEEEtran.bst: No hyphenation pattern has been}%
\typeout{** loaded for the language `#1'. Using the pattern for}%
\typeout{** the default language instead.}%
\else
\language=\csname l@#1\endcsname
\fi
#2}}
\providecommand{\BIBdecl}{\relax}
\BIBdecl

\bibitem{lu2016land}
M.~Lu, J.~Chen, H.~Tang, Y.~Rao, P.~Yang, and W.~Wu, ``Land cover change
  detection by integrating object-based data blending model of landsat and
  modis,'' \emph{Remote Sensing of Environment}, vol. 184, pp. 374--386, 2016.

\bibitem{zhu2014continuous}
Z.~Zhu and C.~E. Woodcock, ``Continuous change detection and classification of
  land cover using all available landsat data,'' \emph{Remote sensing of
  Environment}, vol. 144, pp. 152--171, 2014.

\bibitem{portillo2012forest}
C.~Portillo-Quintero, A.~Sanchez, C.~Valbuena, Y.~Gonzalez, and J.~Larreal,
  ``Forest cover and deforestation patterns in the northern andes (lake
  maracaibo basin): a synoptic assessment using modis and landsat imagery,''
  \emph{Applied Geography}, vol.~35, no. 1-2, pp. 152--163, 2012.

\bibitem{schultz2016performance}
M.~Schultz, J.~G. Clevers, S.~Carter, J.~Verbesselt, V.~Avitabile, H.~V. Quang,
  and M.~Herold, ``Performance of vegetation indices from landsat time series
  in deforestation monitoring,'' \emph{International journal of applied earth
  observation and geoinformation}, vol.~52, pp. 318--327, 2016.

\bibitem{kim2017mapping}
D.~Kim, H.~Lee, A.~Laraque, R.~M. Tshimanga, T.~Yuan, H.~C. Jung, E.~Beighley,
  and C.-H. Chang, ``Mapping spatio-temporal water level variations over the
  central congo river using palsar scansar and envisat altimetry data,''
  \emph{International Journal of Remote Sensing}, vol.~38, no.~23, pp.
  7021--7040, 2017.

\bibitem{yoon2016estimating}
Y.~Yoon, E.~Beighley, H.~Lee, T.~Pavelsky, and G.~Allen, ``Estimating flood
  discharges in reservoir-regulated river basins by integrating synthetic swot
  satellite observations and hydrologic modeling,'' \emph{Journal of Hydrologic
  Engineering}, vol.~21, no.~4, p. 05015030, 2016.

\bibitem{gholizadeh2016comprehensive}
M.~H. Gholizadeh, A.~M. Melesse, and L.~Reddi, ``A comprehensive review on
  water quality parameters estimation using remote sensing techniques,''
  \emph{Sensors}, vol.~16, no.~8, p. 1298, 2016.

\bibitem{roy2014landsat}
D.~P. Roy, M.~A. Wulder, T.~R. Loveland, C.~E. Woodcock, R.~G. Allen, M.~C.
  Anderson, D.~Helder, J.~R. Irons, D.~M. Johnson, R.~Kennedy \emph{et~al.},
  ``{Landsat-8}: Science and product vision for terrestrial global change
  research,'' \emph{Remote sensing of Environment}, vol. 145, pp. 154--172,
  2014.

\bibitem{yokoya2017HS_MS_fusinoRev}
N.~Yokoya, C.~Grohnfeldt, and J.~Chanussot, ``Hyperspectral and multispectral
  data fusion: A comparative review of the recent literature,'' \emph{IEEE
  Geoscience and Remote Sensing Magazine}, vol.~5, no.~2, pp. 29--56, 2017.

\bibitem{Borsoi_2018_Fusion}
R.~A. {Borsoi}, T.~{Imbiriba}, and J.~C.~M. {Bermudez}, ``Super-resolution for
  hyperspectral and multispectral image fusion accounting for seasonal spectral
  variability,'' \emph{IEEE Transactions on Image Processing}, vol.~29, no.~1,
  pp. 116--127, 2020.

\bibitem{loncan2015pansharpeningReview}
L.~Loncan, L.~B. De~Almeida, J.~M. Bioucas-Dias, X.~Briottet, J.~Chanussot,
  N.~Dobigeon, S.~Fabre, W.~Liao, G.~A. Licciardi, M.~Simoes \emph{et~al.},
  ``Hyperspectral pansharpening: A review,'' \emph{IEEE Geoscience and remote
  sensing magazine}, vol.~3, no.~3, pp. 27--46, 2015.

\bibitem{belgiu2019spatiotemporalFusionRev}
M.~Belgiu and A.~Stein, ``Spatiotemporal image fusion in remote sensing,''
  \emph{Remote sensing}, vol.~11, no.~7, p. 818, 2019.

\bibitem{zhu2018spatiotemporalFusReview}
X.~Zhu, F.~Cai, J.~Tian, and T.~K.-A. Williams, ``Spatiotemporal fusion of
  multisource remote sensing data: Literature survey, taxonomy, principles,
  applications, and future directions,'' \emph{Remote Sensing}, vol.~10, no.~4,
  p. 527, 2018.

\bibitem{gao2015fusingLandsatMODISreview}
F.~Gao, T.~Hilker, X.~Zhu, M.~Anderson, J.~Masek, P.~Wang, and Y.~Yang,
  ``Fusing {Landsat} and {MODIS} data for vegetation monitoring,'' \emph{IEEE
  Geoscience and Remote Sensing Magazine}, vol.~3, no.~3, pp. 47--60, 2015.

\bibitem{gao2006STARFM}
F.~Gao, J.~Masek, M.~Schwaller, and F.~Hall, ``On the blending of the landsat
  and {MODIS} surface reflectance: Predicting daily {Landsat} surface
  reflectance,'' \emph{IEEE Transactions on Geoscience and Remote sensing},
  vol.~44, no.~8, pp. 2207--2218, 2006.

\bibitem{zhu2010fusion_ESTARFM}
X.~Zhu, J.~Chen, F.~Gao, X.~Chen, and J.~G. Masek, ``An enhanced spatial and
  temporal adaptive reflectance fusion model for complex heterogeneous
  regions,'' \emph{Remote Sensing of Environment}, vol. 114, no.~11, pp.
  2610--2623, 2010.

\bibitem{hilker2009STAARCH_fusion}
T.~Hilker, M.~A. Wulder, N.~C. Coops, J.~Linke, G.~McDermid, J.~G. Masek,
  F.~Gao, and J.~C. White, ``A new data fusion model for high spatial-and
  temporal-resolution mapping of forest disturbance based on {Landsat} and
  {MODIS},'' \emph{Remote Sensing of Environment}, vol. 113, no.~8, pp.
  1613--1627, 2009.

\bibitem{keshava2002unmixingReview}
N.~Keshava and J.~F. Mustard, ``Spectral unmixing,'' \emph{IEEE signal
  processing magazine}, vol.~19, no.~1, pp. 44--57, 2002.

\bibitem{li2021AEC_SU_modelbased}
H.~Li, R.~A. Borsoi, T.~Imbiriba, P.~Closas, J.~C. Bermudez, and
  D.~Erdo{\u{g}}mu{\c{s}}, ``Model-based deep autoencoder networks for
  nonlinear hyperspectral unmixing,'' \emph{IEEE Geoscience and Remote Sensing
  Letters}, 2021.

\bibitem{borsoi2020BMUAN}
R.~A. Borsoi, T.~Imbiriba, J.~C.~M. Bermudez, and C.~Richard, ``A blind
  multiscale spatial regularization framework for kernel-based spectral
  unmixing,'' \emph{IEEE Transactions on Image Processing}, vol.~29, pp.
  4965--4979, 2020.

\bibitem{zurita2008unmixingFusion1}
R.~Zurita-Milla, J.~G. Clevers, and M.~E. Schaepman, ``Unmixing-based landsat
  {TM} and {MERIS} {FR} data fusion,'' \emph{IEEE Geoscience and Remote Sensing
  Letters}, vol.~5, no.~3, pp. 453--457, 2008.

\bibitem{amoros2013multitemporalFusionUnmixing}
J.~Amor{\'o}s-L{\'o}pez, L.~G{\'o}mez-Chova, L.~Alonso, L.~Guanter,
  R.~Zurita-Milla, J.~Moreno, and G.~Camps-Valls, ``Multitemporal fusion of
  {Landsat/TM} and {ENVISAT/MERIS} for crop monitoring,'' \emph{International
  journal of Applied earth observation and Geoinformation}, vol.~23, pp.
  132--141, 2013.

\bibitem{wu2012unmixingFusion2}
M.~Wu, Z.~Niu, C.~Wang, C.~Wu, and L.~Wang, ``Use of {MODIS} and {Landsat} time
  series data to generate high-resolution temporal synthetic landsat data using
  a spatial and temporal reflectance fusion model,'' \emph{Journal of Applied
  Remote Sensing}, vol.~6, no.~1, p. 063507, 2012.

\bibitem{borsoi2020variabilityReview}
R.~A. Borsoi, T.~Imbiriba, J.~C.~M. Bermudez, C.~Richard, J.~Chanussot,
  L.~Drumetz, J.-Y. Tourneret, A.~Zare, and C.~Jutten, ``Spectral variability
  in hyperspectral data unmixing: A comprehensive review,'' \emph{IEEE
  Geoscience and Remote Sensing Magazine}, 2021, doi:
  10.1109/MGRS.2021.3071158.

\bibitem{Borsoi_multiscaleVar_2018}
R.~A. {Borsoi}, T.~{Imbiriba}, and J.~C. {Moreira Bermudez}, ``A data dependent
  multiscale model for hyperspectral unmixing with spectral variability,''
  \emph{IEEE Transactions on Image Processing}, vol.~29, pp. 3638--3651, 2020.

\bibitem{liu2019reviewCD}
S.~Liu, D.~Marinelli, L.~Bruzzone, and F.~Bovolo, ``A review of change
  detection in multitemporal hyperspectral images: Current techniques,
  applications, and challenges,'' \emph{IEEE Geoscience and Remote Sensing
  Magazine}, vol.~7, no.~2, pp. 140--158, 2019.

\bibitem{borsoi2021MT_MESMA}
R.~A. Borsoi, T.~Imbiriba, J.~C.~M. Bermudez, and C.~Richard, ``Fast unmixing
  and change detection in multitemporal hyperspectral data,'' \emph{IEEE
  Transactions on Computational Imaging}, vol.~7, pp. 975--988, 2021.

\bibitem{erturk2015sparseSU_CD}
A.~Ert{\"u}rk, M.-D. Iordache, and A.~Plaza, ``Sparse unmixing-based change
  detection for multitemporal hyperspectral images,'' \emph{IEEE Journal of
  Selected Topics in Applied Earth Observations and Remote Sensing}, vol.~9,
  no.~2, pp. 708--719, 2015.

\bibitem{huang2012spatiotemporalFusionDictLearning}
B.~Huang and H.~Song, ``Spatiotemporal reflectance fusion via sparse
  representation,'' \emph{IEEE Transactions on Geoscience and Remote Sensing},
  vol.~50, no.~10, pp. 3707--3716, 2012.

\bibitem{borsoi2018superpixels1_sparseU}
R.~A. {Borsoi}, T.~{Imbiriba}, J.~C.~M. {Bermudez}, and C.~{Richard}, ``A fast
  multiscale spatial regularization for sparse hyperspectral unmixing,''
  \emph{IEEE Geoscience and Remote Sensing Letters}, vol.~16, no.~4, pp.
  598--602, April 2019.

\bibitem{song2018spatiotemporalFusionCNNs}
H.~Song, Q.~Liu, G.~Wang, R.~Hang, and B.~Huang, ``Spatiotemporal satellite
  image fusion using deep convolutional neural networks,'' \emph{IEEE Journal
  of Selected Topics in Applied Earth Observations and Remote Sensing},
  vol.~11, no.~3, pp. 821--829, 2018.

\bibitem{shen2016integratedSpatioTemporalFusion}
H.~Shen, X.~Meng, and L.~Zhang, ``An integrated framework for the
  spatio--temporal--spectral fusion of remote sensing images,'' \emph{IEEE
  Transactions on Geoscience and Remote Sensing}, vol.~54, no.~12, pp.
  7135--7148, 2016.

\bibitem{huang2013spatiotemporalFusionBayes}
B.~Huang, H.~Zhang, H.~Song, J.~Wang, and C.~Song, ``Unified fusion of
  remote-sensing imagery: Generating simultaneously high-resolution synthetic
  spatial--temporal--spectral earth observations,'' \emph{Remote sensing
  letters}, vol.~4, no.~6, pp. 561--569, 2013.

\bibitem{xue2017bayesianFusionPixelCovariances}
J.~Xue, Y.~Leung, and T.~Fung, ``A bayesian data fusion approach to
  spatio-temporal fusion of remotely sensed images,'' \emph{Remote Sensing},
  vol.~9, no.~12, p. 1310, 2017.

\bibitem{zhou2020kalmanFusionLandsatMODIS}
F.~Zhou and D.~Zhong, ``Kalman filter method for generating time-series
  synthetic landsat images and their uncertainty from {Landsat} and {MODIS}
  observations,'' \emph{Remote Sensing of Environment}, vol. 239, p. 111628,
  2020.

\bibitem{sedano2014kalmanFUsionNDVI}
F.~Sedano, P.~Kempeneers, and G.~Hurtt, ``A {Kalman} filter-based method to
  generate continuous time series of medium-resolution {NDVI} images,''
  \emph{Remote Sensing}, vol.~6, no.~12, pp. 12\,381--12\,408, 2014.

\bibitem{sarkka2013bayesianBook}
S.~S{\"a}rkk{\"a}, \emph{Bayesian filtering and smoothing}.\hskip 1em plus
  0.5em minus 0.4em\relax Cambridge University Press, 2013, no.~3.

\bibitem{borsoi2020multitemporalUKalmanEM}
R.~A. Borsoi, T.~Imbiriba, P.~Closas, J.~C.~M. Bermudez, and C.~Richard,
  ``Kalman filtering and expectation maximization for multitemporal spectral
  unmixing,'' \emph{IEEE Geoscience and Remote Sensing Letters}, 2020.

\bibitem{mehra1972approaches}
R.~Mehra, ``Approaches to adaptive filtering,'' \emph{IEEE Transactions on
  automatic control}, vol.~17, no.~5, pp. 693--698, 1972.

\bibitem{kitagawa1987smoother}
G.~Kitagawa, ``{Non-Gaussian} state-space modeling of nonstationary time
  series,'' \emph{Journal of the American statistical association}, vol.~82,
  no. 400, pp. 1032--1041, 1987.

\bibitem{gao1996ndwi}
B.-C. Gao, ``{NDWI}--a normalized difference water index for remote sensing of
  vegetation liquid water from space,'' \emph{Remote sensing of environment},
  vol.~58, no.~3, pp. 257--266, 1996.

\bibitem{Vermote_modis_2015}
E.~F. Vermote, J.~C. Roger, and J.~P. Ray, ``{MODIS Surface Reflectance
  User’s Guide},'' NASA, Tech. Rep., May 2015.

\end{thebibliography}




\thebiography

\begin{biographywithpic}
{Haoqing Li}{figures/Haoqing}
received the B.S. degree in electrical
engineering from Wuhan University, China, in 2016
and the M.S. degree in electrical and computer engineering from Northeastern University, Boston, MA, in 2018, where he is currently working toward the
Ph.D. degree in electrical and computer engineering.
His research interests include GNSS signal processing, anti-jamming technology, and robust statistics.
\end{biographywithpic} 

\begin{biographywithpic}
{Bhavya Duvvuri}{figures/Bhavya_2}
received B.Tech. in Biotechnology from Visvesvaraya technological university, Belgaum, India in 2013 and M.S. in Civil and Environmental Engineering from Carnegie Mellon University, Pittsburgh in 2019. She is currently working toward Ph.D. degree in interdisciplinary engineering and her research interests include hydrological modeling, remote sensing, machine learning. 
\end{biographywithpic} 

\begin{biographywithpic}
{Ricardo Borsoi}{figures/Photo_Borsoi}
received the MSc degree in electrical engineering from Federal University of Santa Catarina (UFSC), Florian\'opolis, Brazil, in 2016, and the doctoral degree in sciences pour l'ingenieur and electrical engineering, from Universit\'e C\^ote d'Azur (UCA) and from UFSC, in 2021. His research interests include signal and image processing, machine learning, tensor decomposition, and hyperspectral image analysis.
\end{biographywithpic}

\begin{biographywithpic}
{Tales Imbiriba}{figures/Photo_Imbiriba}
is a research scientist associate with the Electrical and Computer Engineering Department at Northeastern University. He received his B.Sc. (2006) and M.Sc. (2008) degrees both in EE from the Federal University of Para (Brazil), and Ph.D. degree in ECE (2016) from the Federal University of Santa Catarina (Brazil). His research interests focus on statistical signal processing and machine learning, specially regarding problems related to image fusion, spectral unmixing, kernel methods and Bayesian filtering.
\end{biographywithpic} 

\begin{biographywithpic}
{Edward Beighley}{figures/beighley-e_2}
is a Professor in the Department of Civil and Engineering at Northeastern University. He is also an affiliated faculty member in the Global Resilience Institute and the Department of Marine and Environmental Sciences. His research integrates satellite remote sensing and numerical modeling to characterize hydrologic hazards and risks for current and future climate and land use conditions.  His use of remote sensing enables the development of novel applications that support the design and management of civil infrastructure in developing regions where in-situ data are often limited.  Beighley’s research builds on his experience working in the insurance industry, where he served as the technical lead for hydrological science research at FM Global, a worldwide leader in commercial and industrial property insurance.  He blends both his academic and industrial experience to develop novel applications to enable sustainable and resilient communities.  His research portfolio is diverse with funding from national, state and local agencies, private industry and non-profit organizations. He received the National Air and Space Administration’s prestigious New Investigator Award and his most recent projects include National Geographic, NSF Hydrologic Sciences and NASA grants supporting the Surface and Ocean Water Topography (SWOT) and the Gravity Recovery and Climate Experiment Follow-on (GRACE-FO) missions.

\end{biographywithpic} 

\begin{biographywithpic}
{Deniz Erdogmus}{figures/Deniz} received BS in EE and Mathematics (1997), and MS in EE (1999) from the Middle East Technical University, PhD in ECE (2002) from the University of Florida, where he was a postdoc until 2004. He was with CSEE and BME Departments at OHSU (2004-2008). Since 2008, he has been with the ECE Department at Northeastern University. His research focuses on statistical signal processing and machine learning with applications data analysis, human-cyber-physical systems, sensor fusion and intent inference for autonomy. He has served as associate editor and technical committee member for various journals and conferences.
\end{biographywithpic} 

\begin{biographywithpic}
{Pau Closas}{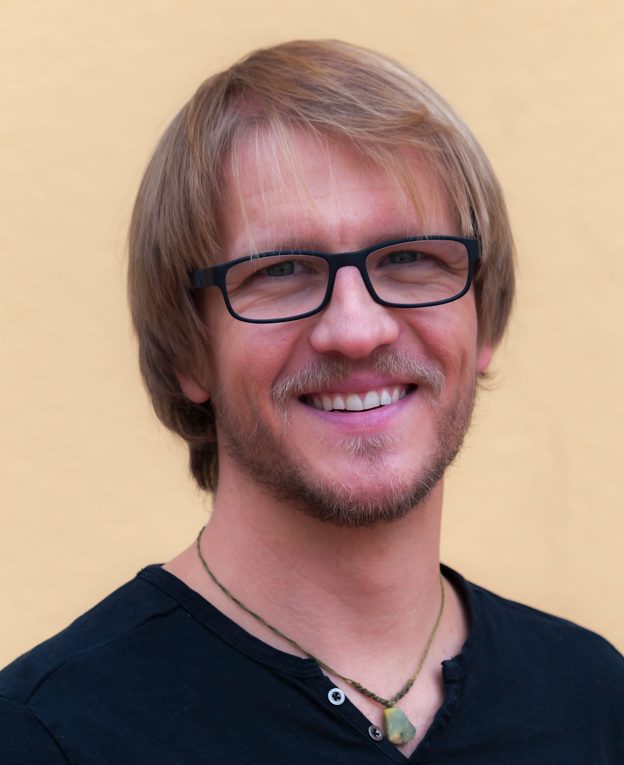}
(S'04 -- M'10 -- SM'13) is an Assistant Professor in Electrical and Computer Engineering at Northeastern University, Boston MA. He received the MS and PhD in Electrical Engineering from UPC in 2003 and 2009, respectively. He also holds a MS in Advanced Maths and Mathematical Engineering from UPC since 2014. He is the recipient of the EURASIP Best PhD Thesis Award 2014, the 9th Duran Farell Award for Technology Research, the 2016 ION's Early Achievements Award, and a 2019 NSF CAREER Award. His primary areas of interest include statistical signal processing, stochastic filtering, robust filtering, and machine learning, with main applications to positioning and localization systems. He volunteered in editorial roles (e.g. NAVIGATION, Proc. IEEE, IEEE Trans. Veh. Tech., and IEEE Sig. Process. Mag.), and has been actively involved in organizing committees of a number of conference such as EUSIPCO (2011, 2019-2022), IEEE SSP'16, IEEE/ION PLANS'20, or IEEE ICASSP'20.
\end{biographywithpic}


\end{document}